\documentclass[12pt]{article}
\usepackage{epsfig,amsmath,amsfonts,amssymb,makeidx,ifthen}
\makeatletter
\@addtoreset{equation}{section}
\makeatother

\makeatletter
\long\def\@makecaption#1#2{{\small
\advance\leftskip1cm
\advance\rightskip1cm
\vskip\abovecaptionskip
\sbox\@tempboxa{#1: #2}%
\ifdim \wd\@tempboxa >\hsize
 #1: #2\par
\else
\global \@minipagefalse
\hb@xt@\hsize{\hfil\box\@tempboxa\hfil}%
\fi
\vskip\belowcaptionskip}}
\makeatother
\def\eq#1\en{\begin{equation}#1\end{equation}}  
\def\eqa#1\ena{\begin{align}#1\end{align}}
\def\eqg#1\eng{\begin{gather}#1\end{gather}}
\newcommand{\lb}[1]{\label{e:#1}}
\newcommand{\rlb}[1]{\eqref{e:#1}} 
\newcommand{\nl}{\notag\\}

\newcommand{\norm}[1]{\left\Vert#1\right\Vert}

\newcommand{\sbkt}[1]{\langle#1\rangle}
\newcommand{\bbkt}[1]{\bigl\langle#1\bigr\rangle}
\newcommand{\Bbkt}[1]{\Bigl\langle#1\Bigr\rangle}

\newcommand{\Hilb}{\mathcal{H}}
\newcommand{\HiS}{\Hilb_\mathrm{S}}
\newcommand{\HiB}{\Hilb_\mathrm{B}}
\newcommand{\HS}{H_\mathrm{S}}
\newcommand{\HB}{H_\mathrm{B}}
\newcommand{\jvec}{\vec{j}}
\newcommand{\betavec}{\vec{\beta}}
\newcommand{\Ejvec}{\vec{E}_{\vec{j}}}
\newcommand{\Ejvecp}{\vec{E}_{\vec{j}'}}
\newcommand{\rhoBb}{\rho_{\mathrm{B},\betavec}}
\newcommand{\rhoss}{\rho_\mathrm{ss}}
\newcommand{\rhosst}{\tilde{\rho}_\mathrm{ss}}
\newcommand{\Tr}{\operatorname{Tr}}
\newcommand{\TrB}{\Tr_\mathrm{B}}
\newcommand{\TrS}{\Tr_\mathrm{S}}
\newcommand{\Ssym}{S_\mathrm{sym}}
\newcommand{\TR}{\mathcal{T}}
\newcommand{\bsr}{\boldsymbol{r}}
\newcommand{\bsp}{\boldsymbol{p}}
\newcommand{\bsone}{\boldsymbol{1}}
\newcommand{\bsoneB}{\bsone_\mathrm{B}}
\newcommand{\bsoneS}{\bsone_\mathrm{S}}
\newcommand{\tiL}{\tilde{L}}
\newcommand{\calN}{\mathcal{N}}
\newcommand{\oeo}{O(\epsilon)}
\newcommand{\oes}{O(\epsilon^2)}

\newcommand{\odo}{O(\delta)}
\newcommand{\ods}{O(\delta^2)}
\newcommand{\oeod}{O(\epsilon\delta)}
\newcommand{\oeods}{O(\epsilon\delta^2)}
\newcommand{\oesd}{O(\epsilon^2\delta)}

\begin{document}
\noindent
{\bf\large Extended Clausius Relation and Entropy for Nonequilibrium Steady States in Heat Conducting Quantum Systems}
\par\bigskip

\noindent
{\small Keiji Saito\footnote{
Graduate School of Science, University of Tokyo, Hongo, Bunkyo-ku, Tokyo 113-0033, Japan.
} and Hal Tasaki\footnote{
Department of Physics, Gakushuin University, Mejiro, Toshima-ku, Tokyo 171-8588,
 Japan.
}}

\begin{abstract}
Recently, in their attempt to construct steady state thermodynamics (SST), Komatsu, Nakagwa, Sasa, and Tasaki found an extension of the Clausius relation to nonequilibrium steady states in classical stochastic processes.
Here we derive a quantum mechanical version of the extended Clausius relation.
We consider a small system of interest attached to large systems which play the role of heat baths.
By only using the genuine quantum dynamics, we realize a heat conducting nonequilibrium steady state in the small system.
We study the response of the steady state when the parameters of the system are changed abruptly, and show that the extended Clausius relation, in which ``heat'' is replaced by the ``excess heat'', is valid when the temperature difference is small.
Moreover we show that the entropy that appears in the relation is similar to  von Neumann entropy but has an extra symmetrization with respect to time-reversal.
We believe that the present work opens a new possibility in the study of nonequilibrium phenomena in quantum systems, and also confirms the robustness of the approach by Komtatsu et al.

\end{abstract}

\section{Introduction}
To understand universal aspects of nonequilibrium states in macroscopic systems, and develop nonequilibrium thermodynamics and statistical mechanics is a remaining big challenge in modern physics.
Nonequilibrium steady states (NESS), which have no macroscopic changes in time but have nonvanishing macroscopic flows of energy and/or matter, may be the best candidate for developing such theories.
In fact there has been some remarkable progress in the understanding of NESS, which includes linear response theory \cite{LR}, formal representation of the probability distribution or density matrix of NESS \cite{Zubarev,McLennan,KG}, fluctuation theorem \cite{Evans,GC,Crooks,Kurchan,Tasaki,hanggi},
 and explicit computation of large deviation functions in specific solvable models \cite{LD1,LD2}.
In spite of all these, however, we are still very much far from the goal of constructing statistical mechanics of NESS.

Recently, Komatsu, Nakagawa, Sasa, and Tasaki \cite{KNST1,KNST2}, who were inspired by earlier works by Oono and Paniconi and \cite{OP} by Landauer \cite{Landauer}, proposed a new direction in thermodynamics for NESS or steady state thermodynamics (SST).
They concentrated on the Clausius relation in the conventional operational thermodynamics, and looked for its extension to NESS.
In a large class of classical many-body systems described by stochastic dynamics, they derived a natural extension of the Clausius relation when  the ``degree of nonequilibrium'' is small.
In their extended Clausius relation, heat in the original relation is replaced by a ``renormalized'' quantity called excess heat\footnote{
The same result was obtained earlier by Ruelle for models with Gaussian thermostat \cite{Ruelle}.
}.
Moreover they found that the entropy which appears in the extended Clausius relation has a very suggestive microscopic expression, which is similar to the Shannon entropy but has an extra symmetrization with respect to time-reversal.
They also found a ``nonlinear nonequilibrium'' improvement of the extended Clausius relation, which involves a correlation between heat and work \cite{KNST2}.
Although we still do not know whether the work of Komatsu et al. leads to a truly nontrivial and powerful applications, it seems fair to say that their work has revealed the existence of a rich nontrivial structure of operational thermodynamics in NESS at least when the ``degree of nonequilibrium'' is sufficiently small.

The purpose of the present work is to extend the approach by Komatsu et al. to quantum systems.
We study a quantum mechanical system which consists of a small system of interest and several large heat baths attached to it which have different temperatures.
By letting the whole system evolve according to the standard quantum dynamics, we realize a heat conducting NESS in the small system.
Then we slightly change the model parameters (such as the temperatures of the baths) and examine the resulting change in the state of the system and the associated flows of heat.
We show that the extended Clausius relation completely analogous to the classical one holds in this situation.
Moreover the entropy that enters the relation has a microscopic expression which is similar to  the von Neumann entropy but with an extra symmetrization with respect to time-reversal.

Our derivation here is a quantum analogue of the derivation of the (classical) extended Clausius relation found in \cite{KNST2}, which is considerably simpler than the original derivation in \cite{KNST1}.
Since the derivation in \cite{KNST2} relies on the linear response representation of the probability distribution in NESS, we shall here make use of the global time-reversal symmetry to derive the corresponding representation for the density matrix of NESS through a Zubarev-McLennan type representation.
We also develop purely quantum mechanical techniques to control small error terms involved in the extended Clausius relation.
We wish to stress that the possibility of the quantum version had been far from obvious as the derivation of the classical version \cite{KNST1,KNST2,Ruelle} heavily relies on various properties specific to classical (stochastic) systems.
As for the ``nonlinear nonequilibrium'' improvement in \cite{KNST2}, we still have little hope of extending it to quantum systems.

We believe that our quantum generalization of the extended Clausius relation is of importance at least from a theoretical point of view.
Since quantum formulation generally contains richer information than classical one, the success of the generalization shows that the relation is indeed robust.
It is particularly important that we encountered the symmetrized von Neumann entropy, which is the most natural among imaginable quantum analogues of the symmetrized Shannon entropy\footnote{
A classical system is, very roughly speaking, a quantum system without off-diagonal matrix elements.
Thus one can imagine a variety of ``quantum extensions'' when only given the classical formulation \rlb{Ssym0} of the symmetrized entropy.
}.
This fact dispels the doubt that the very suggestive expression of the symmetrized Shannon entropy \rlb{Ssym0} might be a mere accident which does not reflect anything deeper.
We may conclude that our quantum generalization strengthens our faith on the universality of the extended Clausius relation and the associated symmetrized (Shannon or von Neumann) entropy.

\bigskip

The present paper is organized as follows.
In the rest of the present section, we give a very brief account of the basic idea and result of Komatsu et al.
In section~\ref{s:Setting}, we define the problem we study and state our main result.
In section~\ref{s:derivation}, we carefully describe our derivation, dividing it into two parts.
In section~\ref{s:dicsussions}, we summarize our findings, and discuss some remaining issues.

\subsubsection*{Clausius relation and extended Clausius relation}
Before discussing the quantum extension, we shall present a very brief review of the extended Clausius relation of Komatsu, Nakagawa, Sasa, and Tasaki\footnote{
We recommend the introductory parts of \cite{KNST2} for a more careful overview of their work.
}.

\begin{figure}[btp]
\begin{center}
\includegraphics[width=12cm]{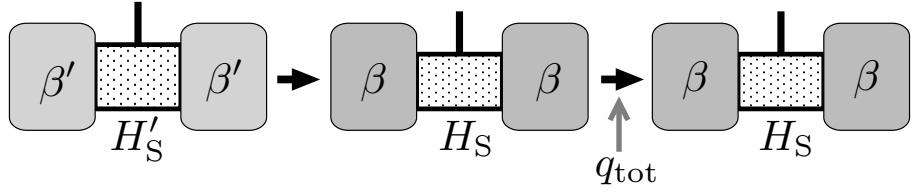}
\end{center}
\caption[dummy]{
Thermodynamic operation in an equilibrium state.
Left: One starts from the equilibrium state with the inverse temperature $\beta'$.
Middle: One abruptly changes the inverse temperature of the baths and the Hamiltonian.  The system is no longer in equilibrium.
Right:  After a sufficiently long time, the system relaxes to a new equilibrium.
The heat $q_\mathrm{tot}$ absorbed by the system during the relaxation plays an essential role in the Clausius relation \rlb{Cl}.
}
\label{fig:OPEQ}
\end{figure}

Let us start from the Clausius relation in the conventional thermodynamics (Figure~\ref{fig:OPEQ}).
Suppose that there is a system governed by the Hamiltonian $\HS'$ and attached to a heat bath with the inverse temperature $\beta'$.
After a sufficiently long time, the system settles to the equilibrium state determined by $\beta'$ and $\HS'$.
Then one makes a thermodynamic operation\footnote{
In the present paper, we put primes on the quantities before the operation, for notational convenience.
We note that the opposite notation is used in \cite{KNST2}.
} by abruptly changing the inverse temperature of the bath from $\beta'$ to $\beta$, and the Hamiltonian from $\HS'$ to $\HS$.
The change of Hamiltonian is a theoretical implementation of various mechanical operation (such as the change of volume) on the system.

The system is no longer in equilibrium after the operation, but will eventually relax to a new equilibrium characterized by $\beta$ and $\HS$.
Let $q_\mathrm{tot}$ be the total amount of heat that flowed into the system from the heat bath during the relaxation process.
Then the celebrated Clausius relation, which indeed was the birth place of the notion of entropy, reads
\eq
S^\mathrm{(final)}-S^\mathrm{(initial)}=\beta\,q_\mathrm{tot}+\ods,
\lb{Cl}
\en
where the parameter $\delta$ characterizes the amount of the change made in the operation.
Here $S^\mathrm{(final)}$ and $S^\mathrm{(initial)}$ are the entropies of the system long after the operation and before the operation, respectively.
It is also essential that the entropies coincide with the Shannon  entropy of the respective equilibrium probability distributions.

\begin{figure}[btp]
\begin{center}
\includegraphics[width=12cm]{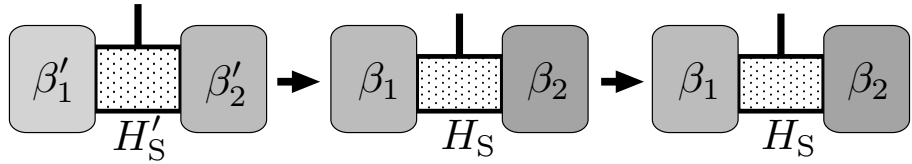}
\end{center}
\caption[dummy]{
Thermodynamic operation in a NESS.
Left: One starts from the NESS  with $\beta'_1$, $\beta'_2$.
Middle: One abruptly changes the inverse temperatures and the Hamiltonian.  The system is no longer in a steady state.
Right:  After a sufficiently long time, the system relaxes to a new NESS.
In the extended Clausius relation \rlb{exCl}, the excess heat or the ``renormalized'' heat plays an essential role since the ``bare'' heat diverges in time.
}
\label{fig:OPNESS}
\end{figure}

Let us now consider the nonequilibrium case (Figure~\ref{fig:OPNESS}).
Suppose that the same system with Hamiltonian $\HS'$ is now attached to $n$ heat baths with different inverse temperatures $\beta'_1,\ldots,\beta'_n$.
The system can never reach an equilibrium, but is expected to  settle to a nonequilibrium steady state (NESS), which shows no macroscopic changes but has a nonvanishing constant heat flow going through the system.
After the system has settled to the NESS characterized by $\beta'_1,\ldots,\beta'_n$ and $\HS'$, one abruptly changes  the inverse temperatures of the baths from $\beta'_1,\ldots,\beta'_n$ to $\beta_1,\ldots,\beta_n$, and the Hamiltonian from $\HS'$ to $\HS$.
After a long time from this operation, the system relaxes to a new NESS characterized by the new parameters.
One then  asks if there can be a relation like the Clausius relation \rlb{Cl} governing such a transition form a NESS to another NESS.

It is apparent that a straightforward generalization of \rlb{Cl} is impossible.
As there is a nonvanishing heat flow in the system, the total heat transfer $q_\mathrm{tot}$ simply diverges in time.
Instead of this ``bare heat'', Komatsu et al. proposed to use a properly ``renormalized'' quantity called the excess heat $q^{(\nu)}_\mathrm{ex}$, which characterizes the intrinsic heat transfer caused by the operation (see \rlb{qex} below).
For a large class of classical stochastic processes, they showed that the extended Clausius relation
\eq
S_\mathrm{sym}^\mathrm{(final)}-S_\mathrm{sym}^\mathrm{(initial)}=\sum_{\nu=1}^n\beta_\nu\,q^{(\nu)}_\mathrm{ex}+\oesd+\ods
\lb{exCl}
\en
holds in the above thermodynamic process.
Here $\varepsilon$ is the ``degree of nonequilibrium'' defined in \rlb{epdef}.
Moreover they have shown that the entropy in the above relation has a microscopic representation in terms of the probability distribution
\eq
S_\mathrm{sym}[p(\cdot)]:=
-\int d\Gamma\,p(\Gamma)\,\frac{1}{2}\bigl\{\log p(\Gamma)+\log p(\tilde{\Gamma})\bigr\},
\lb{Ssym0}
\en
where $\Gamma=(\bsr_1,\ldots,\bsr_N;\bsp_1,\ldots,\bsp_N)$ is a classical state, and $\tilde{\Gamma}=(\bsr_1,\ldots,\bsr_N;-\bsp_1,\ldots,-\bsp_N)$ denotes its time-reversal.
Note that \rlb{Ssym0} is similar to the Shannon entropy, but has an extra symmetrization with respect to time-reversal.
This expression looks quite suggestive since it takes into account quantitatively the breakdown of time-reversal symmetry, which is one of the essential features of a nonequilibrium state.

\section{Setup and Main Results}
\label{s:Setting}
Let us precisely define the problem we study, and state our main result, i.e., the extended Clausius relation \rlb{main} for quantum heat conducting systems.

\subsubsection*{Basic definitions}
We study the situation where a (small) ``system'' of interest is attached to $n$ large heat baths with different temperatures.
We shall model the whole system as a quantum mechanical system which consists of $n+1$ distinct parts.
The first part is the system\footnote{
In what follows ``system'' always means the first part.
The collection of the $n+1$ parts is referred to as the ``whole system''.
} while the latter are the heat baths.
We assume that the different parts do not exchange particles, but 
the system and each heat bath may exchange energy.
The whole system is isolated from the external world.

More precisely we let $\HiS$ and $\HiB^{(1)},\ldots,\HiB^{(n)}$ be the Hilbert spaces of the system and the $n$ heat baths, respectively.
We denote by\footnote{
By $A:=B$ or $B=:A$ we mean that $A$ is defined in terms of $B$.
} $\HiB:=\HiB^{(1)}\otimes\cdots\otimes\HiB^{(n)}$ the Hilbert space of the baths.
The Hilbert space of the whole system is $\Hilb:=\HiS\otimes\HiB$.

The total Hamiltonian is written as
\eq
H:=\HS+\sum_{\nu=1}^n\HB^{(\nu)}+\gamma H_\mathrm{int},
\lb{H}
\en
where $\HS$, which acts on $\HiS$, is the Hamiltonian of the system\footnote{
Rigorously speaking, $\HS$ in \rlb{H} should read $\HS\otimes\bsoneB$, where $\bsoneB$ is the identity on $\HiB$.
We shall mostly omit the identity for simplicity.
}, and $\HB^{(\nu)}$, which acts on $\HiB^{(\nu)}$, is the Hamiltonian of the $\nu$-th bath.
Note that $\HS,\HB^{(1)},\ldots,\HB^{(\nu)}$ all commute with each other.
The interaction between the system and the bath is described by $H_\mathrm{int}$ which acts on the whole $\Hilb$.
We introduced the parameter $\gamma$ to control the strength of the coupling.
All the Hamiltonians are time independent.

Let us denote by $\TR$ the time-reversal operator\footnote{
In the Schr\"{o}dinger representation, one defines $(\TR\varphi)(\bsr_1,\ldots,\bsr_N):=\varphi^*(\bsr_1,\ldots,\bsr_N)$.
}.
The operator $\TR$ is anti-unitary and satisfies $\TR^2=\bsone$.
We assume that all the Hamiltonians are time-reversal invariant in the sense that, e.g., $\TR\HS\TR=\HS$.

We denote by $E^{(\nu)}_j$ and $P^{(\nu)}_j$, where $j=1,2,\ldots$, the eigenvalue and the projection onto the corresponding eigenstate of the Hamiltonian $\HB^{(\nu)}$ for the $\nu$-th bath.
Note that one has $\TR P^{(\nu)}_{j}\TR=P^{(\nu)}_{j}$.
We shall use a convenient vector notation $\jvec:=(j_1,\ldots,j_n)$ where  $j_\nu$ indicates the state of the $\nu$-th bath.
We also write $\Ejvec:=(E^{(1)}_{j_{1}},\ldots,E^{(n)}_{j_{n}})$ for the energy eigenvalues, and $P_{\jvec}:=P^{(1)}_{j_{1}}\otimes\cdots\otimes P^{(n)}_{j_{n}}$ for the projection.

\subsubsection*{Nonequilibrium steady state}
We construct a heat conducting NESS of the system by using the heat baths which are in (near) equilibrium states with different inverse temperatures.

Assume that the $\nu$-th bath is initially in the canonical equilibrium state with inverse temperature $\beta_\nu$.
The whole state (density matrix) of the baths is then given by
\eq
\rhoBb:=\bigotimes_{\nu=1}^n\frac{e^{-\beta_\nu\HB^{(\nu)}}}{Z_\nu(\beta_\nu)}
=\bigotimes_{\nu=1}^n\sum_j\frac{e^{-\beta_\nu E_j^{(\nu)}}}{Z_\nu(\beta_\nu)}\,P^{(\nu)}_j
=\sum_{\jvec}\frac{e^{-\betavec\cdot\Ejvec}}{Z(\betavec)}\,P_{\jvec},
\lb{rhoBb}
\en
where we wrote $\betavec:=(\beta_1,\ldots,\beta_n)$, and $\betavec\cdot\Ejvec=\sum_{\nu=1}^n\beta_\nu E_j^{(\nu)}$.
The partition functions are defined as $Z_\nu(\beta_\nu):=\sum_je^{-\beta_\nu E_j^{(\nu)}}$ and $Z(\betavec):=\prod_{\nu=1}^nZ_\nu(\beta_\nu)$.

The system is initially in an arbitrary state $\rho_0$ (which is a density matrix on $\HiS$).
We let the whole system evolve for a finite time $\tau$  by the dynamics determined by the total Hamiltonian $H$, starting from the initial state $\rho_0\otimes\rhoBb$.
The state won't change considerably if $\tau$ is small.
If $\tau$ is extremely large, on the other hand, the whole system is expected to reach an equilibrium state with a uniform temperature.
We expect that there exists a range of intermediate values of $\tau$ which is ``long enough'' for the system but is ``short enough'' for the baths.
For such $\tau$, the states of the baths are macroscopically the same as the initial states and hence in the equilibrium with the given inverse temperature $\beta_\nu$, while the system totally forgets about its initial state $\rho_0$ and settles to a unique NESS determined by the inverse temperatures $\betavec$ of the baths.
This is our basic and essential assumption.
Although we are totally unable to prove this statement, it seems to be a natural assumption since nonequilibrium steady states in reality is essentially constructed in such a manner.

Precisely speaking, we define the nonequilibrium steady state of the system by
\eq
\rhoss:=\TrB[\,U\,(\rho_0\otimes\rhoBb)\,U^\dagger\,],
\lb{rhoss}
\en
under the natural (and essential) assumption that the right-hand side is independent of $\rho_0$.
Here $\TrB$ stands for the trace over $\HiB$, and the time evolution operator is given by $U:=\exp[-iH\tau]$, where we have set $\hbar=1$.
Of course $\tau$ is chosen from the  ``intermediate'' range that we discussed above.

\subsubsection*{Operation and extended Clausius relation}
Let us now consider an abrupt change of the parameters, which corresponds to a thermodynamic operation.

Before the operation, the Hamiltonian of the system is $\HS'$ and the inverse temperatures of the baths are set to $\betavec'=(\beta_1',\ldots,\beta_n')$.
We assume that the system has settled to the state $\rhoss'$, which is the  NESS   corresponding to this stetting.

Then at $t=0$, we change the Hamiltonian of the system from $\HS'$ to $\HS$ and the inverse temperatures of the baths from $\betavec'$ to $\betavec$.
We introduce a dimensionless quantity $\delta$ which measure the amount of the change.
To be precise, one can set
\eq
\delta:=\frac{\norm{\HS-\HS'}}{\varepsilon_\text{ref}}
+\sum_{\nu=1}^n\frac{|\beta_\nu-\beta'_\nu|}{\beta'_\nu},
\en
where $\norm{\cdot}$ indicates a suitbable operator norm, and $\varepsilon_\text{ref}$ is a (rather arbitrary) reference energy scale.
The Hamiltonians of the baths and the interaction Hamiltonian are not  changed throughout this consideration.

To implement the change of parameters theoretically, we first prepare the NESS $\rhoss'$ (by using \rlb{rhoss} with replacing $\HS$ with $\HS'$, and $\betavec$ with $\betavec'$) and then couple it with a ``fresh'' equilibrium states $\rhoBb$ of the baths.
Thus the whole system re-starts from the state $\rhoss'\otimes\rhoBb$ and evolves by the dynamics determined by $H$.
This may not be exactly what one does in an experiment, but our derivation works precisely in this situation\footnote{
\label{fn:baths}
In other words, we need to assume certain ``Markovness'' of the heat baths.
Therefore even in the special case where one only changes $\HS$ and keeps $\betavec$ constant, we have to ``refresh'' the baths for a technical reason.
Note that the Markovness of the bath was also assumed in the classical case \cite{KNST1,KNST2}.
}.

Let us concentrate on the flow of heat in the time interval $0\le t\le\tau$ after the parameter change.
We define the total heat that flows into the system from the $\nu$-th bath as the difference in the expectation values of the Hamiltonian of the bath
\eq
q_\mathrm{tot}^{(\nu)}:=\TrB[\,\HB^{(\nu)}\,\rhoBb\,]
-\Tr[\,\HB^{(\nu)}\,U\,(\rhoss'\otimes\rhoBb)\,U^\dagger\,]
=\Tr[\,Q^{(\nu)}\,(\rhoss'\otimes\rhoBb)\,],
\lb{qtot}
\en
where $\Tr$ denotes the trace over the whole Hilbert space $\Hilb$.
Here we defined the heat operator by
\eq
Q^{(\nu)}:=\HB^{(\nu)}-U^\dagger\,\HB^{(\nu)}\,U.
\lb{Qnu}
\en
Note that one can write $Q^{(\nu)}=\int_0^\tau dt\,J^{(\nu)}(t)$ with the current operator $J^{(\nu)}(t):=-i\,e^{iHt}\,[H,\HB^{(\nu)}]\,e^{-iHt}$.

Since there is a steady nonvanishing flow of heat in the NESS, the total heat $q_\mathrm{tot}^{(\nu)}$ generally contains a part which grows proportionally to the total time $\tau$.
To ``renormalize'' this divergence, we consider the heat transfer necessary to maintain the NESS, which is given by
\eq
q_\mathrm{hk}^{(\nu)}:=
\TrB[\,\HB^{(\nu)}\,\rhoBb\,]
-\Tr[\,\HB^{(\nu)}\,U\,(\rhoss\otimes\rhoBb)\,U^\dagger\,]
=\Tr[\,Q^{(\nu)}\,(\rhoss\otimes\rhoBb)\,].
\lb{qhk}
\en
Note that we have simply replaced $\rhoss'$ in \rlb{qtot} with the ``right'' NESS $\rhoss$.
We call $q_\mathrm{hk}^{(\nu)}$ the house-keeping heat after \cite{OP}.
While both $q_\mathrm{tot}^{(\nu)}$ and $q_\mathrm{hk}^{(\nu)}$ grow proportionally to $\tau$, their difference
\eq
q_\mathrm{ex}^{(\nu)}:=q_\mathrm{tot}^{(\nu)}-q_\mathrm{hk}^{(\nu)}
=\Tr[\,Q^{(\nu)}\,(\rhoss'\otimes\rhoBb-\rhoss\otimes\rhoBb)\,],
\lb{qex}
\en
which we call the {\em excess heat}\/, is expected to be independent of $\tau$, and characterize the intrinsic heat transfer caused by the change of parameters.
This definition is of course a natural quantum analogue of the corresponding definition in classical systems \cite{OP,KNST1,KNST2}.

The main result of the present paper is the extended Clausius relation
\eq
\Ssym[\rhoss]-\Ssym[\rhoss']=\sum_{\nu=1}^n\beta_\nu\,q_\mathrm{ex}^{(\nu)}
+\oesd+\ods,
\lb{main}
\en
which holds in the weak coupling limit\footnote{
If one wants to see how the (small) $\gamma$ enters the relation, it suffices to replace $\oesd$ with $\{\oeo+O(\gamma)\}^2\,\odo$.
} with negligible coupling $\gamma$.
The quantity $\epsilon$ is a dimensionless measure of the ``degree of nonequilibrium'', and is defined as
\eq
\epsilon:=
\sum_{\nu=1}^n
\Bigl\{\frac{|\beta_\nu-\beta|}{\beta}
+\frac{|\beta'_\nu-\beta'|}{\beta'}\Bigr\},
\lb{epdef}
\en
where $\beta:=\sum_{\nu=1}^n\beta_\nu/n$ and $\beta':=\sum_{\nu=1}^n\beta'_\nu/n$.
The symmetrized von Neumann entropy, which is the central quantity in the present work, is defined by
\eq
\Ssym[\rho]:=-\TrS\Bigl[\,\rho\,\frac{\log\rho+\log(\TR\rho\TR)}{2}\,\Bigr],
\lb{Ssym}
\en
where $\TrS$ denotes the trace in $\HiS$.

Note that \rlb{main} has precisely the same form as the extended Clausius relation \rlb{exCl} for classical systems.
The quantity $-\sum_{\nu=1}^n\beta_\nu\,q_\mathrm{ex}^{(\nu)}
$ can be interpreted as the {\em excess entropy production}\/ in the heat baths as in the classical case \cite{KNST1,KNST2}.
As mentioned in the introduction, a remarkable fact is that the symmetrized von Neumann entropy \rlb{Ssym} seems to be the most beautiful and (mathematically) natural among imaginable quantum extensions of the symmetrized Shannon entropy \rlb{Ssym0} discovered in \cite{KNST1,KNST2}.

It is natural to ask whether our new result predict any phenomena intrinsic to quantum systems, which cannot be observed in the classical counterpart.
Unfortunately we have so far no idea about experimental settings in which quantum nature of NESS can be observed.
But since our result itself fully reflects quantum aspects of the problem, there is a chance that one can find a truly quantum phenomena described by it.

\section{Derivation}
\label{s:derivation}

We here derive the extended Clausius relation \rlb{main}.
Since the derivation is involved, we divide it into two parts.

In the first part, we derive a representation \rlb{rep} for the matrix elements of $\rhoss$.
The representation is a consequence of the Zubarev-McLennan representation \rlb{Zubarev}, and is closely related to the linear response formula \rlb{LR}.

In the second part, we evaluate the difference $\Ssym[\rhoss]-\Ssym[\rhoss']$ of the symmetrized von Neumann entropy, and use the representation \rlb{rep} to show the desired relation \rlb{main}.

\subsection{Representations of the NESS}
\label{s:representation}
Let us derive some formal representations for the density matrix of the NESS.
The time-reversal symmetry of the whole system plays an essential role here.

Let $\rho_0$ and $\rho_1$ be arbitrary states on $\HiS$.
Take an arbitrary function $f(\jvec,\jvec')$ of the two sets of indices $\jvec=(j_1,\ldots,j_n)$, $\jvec'=(j'_1,\ldots,j'_n)$ for the energy eigenstates of the baths.
We then define the average of $f(\jvec,\jvec')$ by
\eq
\bbkt{f(\jvec,\jvec')}_{\rho_0\to\rho_1}:=
\frac{\sum_{\jvec,\jvec'}f(\jvec,\jvec')\,
\Tr[\,(\rho_1\otimes P_{\jvec'})\,U\,(\rho_0\otimes P_{\jvec})\,U^\dagger\,]
\,e^{-\betavec\cdot\Ejvec}/Z(\betavec)}
{\Tr[\,(\rho_1\otimes\bsoneB)\,U\,(\rho_0\otimes\rhoBb)\,U^\dagger\,]}.
\lb{fav}
\en
The normalization $\sbkt{1}_{\rho_0\to\rho_1}=1$ is easily checked by recalling the definition \rlb{rhoBb} of $\rhoBb$, and noting that $\sum_{\jvec'}P_{\jvec'}=\bsoneB$.
Note also that the denominator can be written as $\TrS[\rho_1\,\rhoss]$ from \rlb{rhoss}, and is independent of $\rho_0$.

The average \rlb{fav} may be interpreted as follows.
The whole system starts from $\rho_0\otimes\rhoBb$, and evolves by the time evolution operator $U$.
At the final moment we test whether the system is in the state $\rho_1$, and leave only those histories in which the result is positive\footnote{
This is a quantum analogue of ``conditioning of final sate'' frequently used in \cite{KNST1,KNST2} and related works.
}.
The denominator is nothing but the probability that we get a positive result in the test.
To get the average, we further make projective measurements of the energies of all the baths at the initial and the final moments to get lists of $\jvec$ and $\jvec'$.

We stress that we are not proposing to execute this highly artificial averaging procedure.
In fact the average \rlb{fav} is used here only as a theoretical tool for deriving the representation \rlb{rep}, which only involves the standard quantum mechanical average.

For an arbitrary state $\rho$, write its time reversal as $\tilde{\rho}:=\TR\rho\TR$.
We will show later that, for any states $\rho$ and $\xi$ of the system, one has
\eq
\frac{\TrS[\,\rho\,\rhosst\,]}{\TrS[\,\xi\,\rhosst\,]}
=
\Bbkt{e^{\betavec\cdot(\Ejvec-\Ejvecp)}}_{\rho\,\to\,\tilde{\xi}}
\lb{Zubarev}
\en
The equality \rlb{Zubarev} is a variant of the Zubarev-McLennan representation \cite{Zubarev,McLennan}, and will be a basis of our study.
If we fix an arbitrary $\xi$ as a reference state, then \rlb{Zubarev} represents the overlap between $\rhosst$ and an arbitrary state $\rho$ in terms of the expectation value related to the transfer of heat bewteen the system and the baths.
See the Appendix for more on the Zubarev-McLennan representation.
We also note in passing that by setting $\xi=\rho$, one gets an interesting nonequilibrium identity
\eq
\Bbkt{e^{\betavec\cdot(\Ejvec-\Ejvecp)}}_{\rho\,\to\,\tilde{\rho}}=1.
\lb{Zubarev1}
\en

When $\rho_0$ and $\rho_1$ are the projections onto pure states $\psi$ and $\varphi$, respectively, the Zubarev-Mclennan representation \rlb{Zubarev} becomes
\eq
\bbkt{\varphi,\rhosst\,\varphi}=\bbkt{\psi,\rhosst\,\psi}\,
\Bbkt{e^{\betavec\cdot(\Ejvec-\Ejvecp)}}_{\varphi\,\to\,\tilde{\psi}}
\lb{Zubarev2}
\en
where $\tilde{\psi}:=\TR\psi$, and $\sbkt{\cdot,\cdot}$ denotes the inner product in $\HiS$.
Here we have made a slight abuse of notation to put the pure states $\varphi$ and $\tilde{\psi}$, rather than the corresponding density matrices, in the subscript.

Suppose further that $\varphi$ is very close to being an eigenstate of $\HS$.
More precisely we assume that there is $\varepsilon$ such that $\sqrt{\sbkt{\varphi,(\HS-\varepsilon)^2\,\varphi}}=:\eta$ is small.
Then we can use the Zubarev-Mclennan representation \rlb{Zubarev2} to derive
\eq
\log\bbkt{\varphi,\rhosst\,\varphi}=
\sum_{\nu=1}^n\beta_\nu\Tr[\,Q^{(\nu)}\,(P_{\varphi}\otimes\rhoBb)\,]
-S+\{O(\eta)+O(\gamma)+\oeo\bigr\}^2,
\lb{rep}
\en
where the heat operator $Q^{(\nu)}$ is defined in \rlb{Qnu}.
Recall that the parameter $\gamma$ introduced in \rlb{H} characterizes the strength of the coupling between the system and the baths.
Here $S$ is a constant independent of $\varphi$, and is given by $S:=-\sum_i\bbkt{\psi_i,\rhoss\,\psi_i}\log\bbkt{\psi_i,\rhoss\,\psi_i}$, where $\psi_i$ is the energy eigenstate of the system, i.e., $\HS\psi_i=\varepsilon_i\psi_i$.

The representation \rlb{rep} of the NESS plays an essential role in our derivation of the extended Clausius relation \rlb{main}.
As we shall discuss in the Appendix, \rlb{rep} is closely related to the linear response formula.

\subsubsection*{Derivation of the Zubarev-McLennan representation \protect{\rlb{Zubarev}}}
We first note that for an arbitrary operator $A$ one has\footnote{
Since $\TR^2=\bsone$, one might erroneously conclude that $\Tr[\TR A\TR]\stackrel{\text{wrong!}}{=}\Tr[\TR^2A]=\Tr[A]$.
One cannot use standard manipulations for an anti-linear operator $\TR$.
}
\eq
\Tr[\TR A\TR]=(\Tr[A])^*.
\lb{TrA}
\en
If $\Tr[A]$ is real, in particular, $\Tr[\TR A\TR]=\Tr[A]$.
We shall make a repeated use of this fact in the following.

Note that 
$\TrS[\rho\rhosst]=\TrS[\tilde{\rho}\rhoss]$
since $\TrS[\rho\rhosst]$ is real.
By using the expression \rlb{rhoss} of the NESS, one observe that
\eqa
\TrS[\tilde{\rho}\rhoss]&=
\Tr[\,(\tilde{\rho}\otimes\bsoneB)\,U\,(\xi\otimes\rhoBb)\,U^\dagger\,]
\nl
&=\sum_{\jvec,\jvec'}
\Tr[\,(\tilde{\rho}\otimes P_{\jvec})\,U\,(\xi\otimes P_{\jvec'})\,U^\dagger\,]\,\frac{e^{-\betavec\cdot\Ejvecp}}{Z},
\notag
\intertext{where we used  the definition \rlb{rhoBb}.
By noting $\TR^2=\bsone$, $\TR P_{\jvec}\TR=P_{\jvec}$, and $\TR U\TR=U^\dagger$, we see that
}
&=\sum_{\jvec,\jvec'}
\Tr[\,\TR\,(\rho\otimes P_{\jvec})\,U^\dagger\,(\tilde{\xi}\otimes P_{\jvec'})\,U\,\TR\,]\,\frac{e^{-\betavec\cdot\Ejvecp}}{Z}
\nl
&=\sum_{\jvec,\jvec'}
\Tr[\,(\rho\otimes P_{\jvec})\,U^\dagger\,(\tilde{\xi}\otimes P_{\jvec'})\,U\,]\,\frac{e^{-\betavec\cdot\Ejvecp}}{Z},
\intertext{where we used the reality of the trace and \rlb{TrA}.
Then from the property of the trace we get}
&=\sum_{\jvec,\jvec'}
\Tr[\,(\tilde{\xi}\otimes P_{\jvec'})\,U\,(\rho\otimes P_{\jvec})\,U^\dagger\,]\,\frac{e^{-\betavec\cdot\Ejvecp}}{Z},
\nl
&=\sum_{\jvec,\jvec'}
e^{\betavec\cdot(\Ejvec-\Ejvecp)}\,
\Tr[\,(\tilde{\xi}\otimes P_{\jvec'})\,U\,(\rho\otimes P_{\jvec})\,U^\dagger\,]\,\frac{e^{-\betavec\cdot\Ejvec}}{Z}
\nl
&=\Bbkt{e^{\betavec\cdot(\Ejvec-\Ejvecp)}}_{\rho\,\to\,\tilde{\xi}}\,
\Tr[\,(\tilde{\xi}\otimes\bsoneB)\,U\,(\rho\otimes\bsoneB)\,U^\dagger\,],
\ena
where we used the definition \rlb{fav}.
By recalling \rlb{rhoss}, we see that this is the desired \rlb{Zubarev}.

\subsubsection*{Derivation of the representation \protect{\rlb{rep}}}
Let $\varphi$ be a state that satisfies the condition for $\rlb{rep}$, and $\psi_i$ be an arbitrary eigenstate of $\HS$.
We first show that
\eq
\bbkt{\varphi,\rhosst\,\varphi}=\bbkt{\psi_i,\rhoss\,\psi_i}
\exp\Bigl[\,\bbkt{\betavec\cdot(\Ejvec-\Ejvecp)}_{\varphi\,\to\,\psi_i}+\bigl\{O(\eta)+O(\gamma)+\oeo\bigr\}^2\,\Bigr].
\lb{preLR}
\en

In the process that defines the average $\sbkt{\cdots}_{\varphi\,\to\,\psi_i}$, the whole system starts from $\varphi\otimes(\text{the eigenstate of the baths Hamiltonian indexed by $\jvec$})$ and ends in 
\newline
$\psi_i\otimes(\text{the eigenstate indexed by $\jvec'$})$.
The energy conservation then reads
\eq
\varepsilon+O(\eta)+\sum_{\nu=1}^nE_{j_\nu}^{(\nu)}=\varepsilon_i+\sum_{\nu=1}^nE_{j'_\nu}^{(\nu)}+O(\gamma),
\lb{encon}
\en
where $O(\gamma)$ represents the effect of the interaction between the system and the bath\footnote{
Note that the error term here does not depend on $\tau$, the time interval.
All that we need is the fact that the eigenstate of the whole Hamiltonian $H$ with energy $E$ is a linear combination of the states characterized by $i$, $\jvec$ such that $\varepsilon_i+\sum_{\nu=1}^nE_{j_\nu}^{(\nu)}=E+O(\gamma)$.
}.
By using the averaged inverse temperature $\beta:=\sum_{\nu=1}^n\beta_\nu/n$, we write
\eq
\betavec\cdot(\Ejvec-\Ejvecp)=
\beta\,(\varepsilon_i-\varepsilon)+\mathcal{F}_{\jvec,\jvec'}
\lb{bEF}
\en
with
\eq
\mathcal{F}_{\jvec,\jvec'}:=
\beta\Bigl\{\sum_{\nu=1}^n(E_{j_\nu}^{(\nu)}-E_{j'_\nu}^{(\nu)})-\varepsilon_i+\varepsilon\Bigr\}
+\sum_{\nu=1}^n(\beta_\nu-\beta)(E_{j_\nu}^{(\nu)}-E_{j'_\nu}^{(\nu)})
=O(\eta)+O(\gamma)+\oeo,
\lb{Fdef}
\en
where we used \rlb{encon} and $\beta_\nu-\beta=\oeo$ to get the final estimate.
Noting that $\varepsilon$ and $\varepsilon_i$ are fixed, one finds
\eqa
\Bbkt{e^{\betavec\cdot(\Ejvec-\Ejvecp)}}_{\varphi\,\to\,\psi_i}
&=
e^{\beta(\varepsilon_i-\varepsilon)}\Bbkt{e^{\mathcal{F}_{\jvec,\jvec'}
}}_{\varphi\,\to\,\psi_i}
\nl
&=
e^{\beta(\varepsilon_i-\varepsilon)+
\sbkt{\mathcal{F}_{\jvec,\jvec'}}_{\varphi\,\to\,\psi_i}+O^2}
=
e^{\sbkt{\betavec\cdot(\Ejvec-\Ejvecp)}_{\varphi\,\to\,\psi_i}+O^2},
\ena
where we wrote $O=O(\eta)+O(\gamma)+\oeo$.
Substituting this into \rlb{Zubarev2} and noting that $\tilde{\psi}_i=\psi_i$ and $\bbkt{\psi_i,\rhosst\,\psi_i}=\bbkt{\psi_i,\rhoss\,\psi_i}$, we get the desired \rlb{preLR}.

Let us proceed to show the final goal \rlb{rep}.
By taking the logarithm of \rlb{preLR}, one has
\eq
\log\sbkt{\varphi,\rhosst\,\varphi}=
\bbkt{\betavec\cdot(\Ejvec-\Ejvecp)}_{\varphi\,\to\,\psi_i}+\log\bbkt{\psi_i,\rhoss\,\psi_i}+O^2,
\lb{logprp}
\en
for any $i$.
Recall here that by \rlb{fav}
\eq
\bbkt{\betavec\cdot(\Ejvec-\Ejvecp)}_{\varphi\,\to\,\psi_i}=
\frac{\sum_{\jvec,\jvec'}\betavec\cdot(\Ejvec-\Ejvecp)\,
\Tr[\,(P_{\psi_i}\otimes P_{\jvec'})\,U\,(P_{\varphi}\otimes P_{\jvec})\,U^\dagger\,]
\,e^{-\betavec\cdot\Ejvec}/Z(\betavec)}
{\sbkt{\psi_i,\rhoss\,\psi_i}},
\lb{FF}
\en
where $P_{\psi_i}$ and $P_{\varphi}$ denote the projections onto $\psi_i$ and $\varphi$, respectively.
Since $\sum_iP_{\psi_i}=\bsoneS$, we see that
\eqa
\sum_i&\sbkt{\psi_i,\rhoss\,\psi_i}\bbkt{\betavec\cdot(\Ejvec-\Ejvecp)}_{\varphi\,\to\,\psi_i}
\nl
=&\,
\sum_{\jvec,\jvec'}\betavec\cdot(\Ejvec-\Ejvecp)\,
\Tr[\,(\bsoneS\otimes P_{\jvec'})\,U\,(P_{\varphi}\otimes P_{\jvec})\,U^\dagger\,]
\,\frac{e^{-\betavec\cdot\Ejvec}}{Z(\betavec)}
\nl
=&\,
\sum_{\jvec,\jvec'}\sum_{\nu=1}^n\beta_\nu
\Tr[\,(\bsoneS\otimes P_{\jvec'})\,U\,\HB^{(\nu)}\,(P_{\varphi}\otimes P_{\jvec})\,U^\dagger\,]
\,\frac{e^{-\betavec\cdot\Ejvec}}{Z(\betavec)}
\nl
&-
\sum_{\jvec,\jvec'}\sum_{\nu=1}^n\beta_\nu
\Tr[\,\HB^{(\nu)}\,(\bsoneS\otimes P_{\jvec'})\,U\,(P_{\varphi}\otimes P_{\jvec})\,U^\dagger\,]
\,\frac{e^{-\betavec\cdot\Ejvec}}{Z(\betavec)}
\nl
=&\,\sum_{\nu=1}^n\beta_\nu\Tr[\,\HB^{(\nu)}\,(P_{\varphi}\otimes\rhoBb)\,]-\sum_{\nu=1}^n\beta_\nu\Tr[\,\HB^{(\nu)}\,U\,(P_{\varphi}\otimes\rhoBb)\,U^\dagger\,]
\nl
=&\,\sum_{\nu=1}^n\beta_\nu\Tr[\,Q^{(\nu)}\,(P_{\varphi}\otimes\rhoBb)\,].
\lb{long}
\ena
We now multiply the both sides of \rlb{logprp} by $\bbkt{\psi_i,\rhoss\,\psi_i}$ and sum over all $i$.
Since $\sum_i\bbkt{\psi_i,\rhoss\,\psi_i}=\TrS[\rhoss]=1$, the above \rlb{long} implies the goal \rlb{rep}.

\subsection{Extended Clausius relation}

We are ready to show the extended Clausius relation \rlb{main}.
By substituting the definition \rlb{Ssym} of the symmetrized von Neumann entropy into the left-hand side of  \rlb{main}, we get
\eq
\Ssym[\rhoss]-\Ssym[\rhoss']
=-\frac{1}{2}\TrS[\,\rhoss\,(\log\rhoss+\log\rhosst)\,]
+\frac{1}{2}\TrS[\,\rhoss'\,(\log\rhoss'+\log\rhosst')\,],
\lb{SS}
\en
where $\rhosst=\TR\rhoss\TR$ and $\rhosst'=\TR\rhoss'\TR$.
Let us reorganize the right-hand side of \rlb{SS}, and write
\eq
\Ssym[\rhoss]-\Ssym[\rhoss']=A_1+\frac{1}{2}(A_2+A_3+A_4),
\lb{SSdec}
\en
with
\begin{gather}
A_1:=\TrS[\,(\rhoss'-\rhoss)\,\log\rhosst],
\lb{A1}
\\
A_2:=\TrS[\,(\rhoss-\rhoss')\,(\log\rhosst-\log\rhoss)\,]
=\TrS[\,(e^L-e^{L'})(\tiL-L)\,],
\\
A_3:=\TrS[\,\rhoss'\,(\log\rhoss'-\log\rhoss)\,]
=\TrS[\,e^{L'}(L'-L)\,],
\\
A_4:=\TrS[\,\rhoss'\,(\log\rhosst'-\log\rhosst)\,]
=\TrS[\,e^{L'}(\tiL'-\tiL)\,],
\end{gather}
where we wrote $L:=\log\rhoss$, $L':=\log\rhoss'$, $\tiL:=\log\rhosst$, and $\tiL':=\log\rhosst'$.

\subsubsection*{Evaluation of small terms}
Let us first show that $A_2+A_3+A_4=\oesd+\ods$ and is small.
In what follows we make a repeated use of the standard perturbation formula
\eq
e^{A+B}=e^A+\int_0^1ds\,e^{(1-s)A}\,B\,e^{sA}+O(\norm{B}^2)
\en
for arbitrary operators $A$ and $B$.
We also note that $L-\tiL=\oeo$, $L'-\tiL'=\oeo$ since the equilibrium state is time-reversal invariant, and $L'-L=\odo$, $\tiL'-\tiL=\odo$ by definition of $\delta$.

We start from $A_3$, which is the easiest.
Since $\TrS[e^L]=\TrS[e^{L'}]=1$, one has
\eqa
0&=\TrS[e^L-e^{L'}]=\TrS[e^{L'+(L-L')}-e^{L'}]
\nl&=
\TrS\Bigl[\,\int_0^1ds\,e^{(1-s)L'}(L-L')e^{sL'}\,\Bigr]+\ods
\nl
&=\TrS[\,e^{L'}(L-L')\,]+\ods,
\ena
which implies that $A_3=\ods$, which is negligible.

As for $A_2$, we observe that
\eqa
A_2&=\TrS[\,(e^L-e^{L'})(\tiL-L)\,]=\TrS[\,(e^{L'+(L-L')}-e^{L'})(\tiL-L)\,]
\nl
&=\TrS\Bigl[\,\int_0^1ds\,e^{(1-s)L'}(L-L')e^{sL'}(\tiL-L)\,\Bigr]+\oeods,
\lb{A2}
\ena
which is overall a quantity of $\oeod$, and is not negligible.
We rewrite $A_4$ in a similar manner to get
\eqa
A_4&=\TrS[\,e^{L'}(\tiL'-\tiL)\,]=\TrS[\,e^{\tiL'+(L'-\tiL')}(\tiL'-\tiL)\,]
\nl
&=\TrS\Bigl[\,\int_0^1ds\,e^{(1-s)\tiL'}(L'-\tiL')e^{s\tiL'}(\tiL'-\tiL)\,\Bigr]
+\TrS[\,e^{\tiL'}(\tiL'-\tiL)\,]+\oesd,
\notag
\intertext{where the second term is nothing but the time-reversal of $-A_3$ and is of $\ods$.
Now by noting that $\tiL'-\tiL=\odo$ and $L'-\tiL'=L-\tiL+\odo$, one can replace $L'-\tiL'$ with $L-\tiL$ by only producing terms of $\ods$ to get}
&=\TrS\Bigl[\,\int_0^1ds\,e^{(1-s)\tiL'}(L-\tiL)e^{s\tiL'}(\tiL'-\tiL)\,\Bigr]+\oesd+\ods.
\intertext{By using \rlb{TrA} and the reality of the trace, we take the time-reversal of everything to get}
&=\TrS\Bigl[\,\int_0^1ds\,e^{(1-s)L'}(\tiL-L)e^{sL'}(L'-L)\,\Bigr]+\oesd+\ods
\nl
&=\TrS\Bigl[\,\int_0^1ds\,e^{(1-s)L'}(L'-L)e^{sL'}(\tiL-L)\,\Bigr]+\oesd+\ods,
\ena
where we have changed the order of the operators inside the trace and made a change of variable $s\to1-s$.
Comparing with \rlb{A2} we see that $A_2+A_4=\oesd+\ods$.

\subsubsection*{Evaluation of the main term}
Let us evaluate $A_1$, which gives the main contribution.
Let $\varphi_r$ with $r=1,2,\ldots$ be the eigenstates of $\rhosst$.
Representing the trace using the basis formed by $\varphi_r$, we have from \rlb{A1} that
\eq
A_1=\sum_r\bigl\{\sbkt{\varphi_r,\rhoss'\,\varphi_r}-
\sbkt{\varphi_r,\rhoss\,\varphi_r}\bigr\}\,\log\sbkt{\varphi_r,\rhosst\,\varphi_r}.
\lb{A11}
\en
Since $\rhosst$ becomes the canonical distribution $e^{-\beta\HS}/Z_\mathrm{S}$ when $\epsilon=0$ and $\gamma=0$ (i.e., the equilibrium and the weak coupling limit) one sees that for each $r$ there is $\varepsilon_r$ such that $\sqrt{\sbkt{\varphi_r,(\HS-\varepsilon_r)^2\,\varphi_r}}=O(\gamma)+\oeo$.
In what follows we assume that $\gamma$ is small enough so that $O(\gamma)$ is always absorbed into $\oeo$.

Then the condition for the representation \rlb{rep} is satisfied, and we have
\eq
\log\bbkt{\varphi_r,\rhosst\,\varphi_r}=
\sum_{\nu=1}^n\beta_\nu\Tr[\,Q^{(\nu)}\,(P_{\varphi_r}\otimes\rhoBb)\,]
-S+\oes.
\lb{rep2}
\en
Since $\rhoss'-\rhoss=\odo$, we get from \rlb{A11} and \rlb{rep2} that
\eqa
A_1&=\sum_r\bigl\{\sbkt{\varphi_r,\rhoss'\,\varphi_r}-
\sbkt{\varphi_r,\rhoss\,\varphi_r}\bigr\}\,
\sum_{\nu=1}^n\beta_\nu\Tr[\,Q^{(\nu)}\,(P_{\varphi_r}\otimes\rhoBb)\,]
+\oesd
\notag
\intertext{where we noted that $S$ is independent of $r$, and $\sum_r\bigl\{\sbkt{\varphi_r,\rhoss'\,\varphi_r}-
\sbkt{\varphi_r,\rhoss\,\varphi_r}\bigr\}=\TrS[\rhoss']-\TrS[\rhoss]=0$.
We can further rewrite this as
}
&=\sum_{\nu=1}^n\beta_\nu \Bigl\{\Tr[\,Q^{(\nu)}\,(\rhoss'\otimes\rhoBb)\,]
-\Tr[\,Q^{(\nu)}\,(\rhoss\otimes\rhoBb)\,]
\Bigr\}+\oesd
\nl
&=\sum_{\nu=1}^n\beta_\nu\,q_\mathrm{ex}^{(\nu)}+\oesd,
\ena
where we used the definition \rlb{qex} of excess heat.
By substituting into \rlb{SSdec} all the estimates, we get the desired extended Clausius relation \rlb{main}.

\section{Discussions and future issues}
\label{s:dicsussions}
We constructed a heat conducting nonequilibrium steady state (NESS) in a quantum system, and made use of the time-reversal symmetry to derive a Zubarev-McLennan type representation for NESS.
The representation was then used to derive the extended Clausius relation.
This is a straightforward and very natural quantum generalization of the extended Clausius relation derived by Komatsu et al. in classical systems \cite{KNST1,KNST2}.

As we have stressed in the introduction, the fact that the new relation contains the most natural quantum extension of the symmetrized Shannon entropy is particularly meaningful.
It may be regarded as an indication that the symmetrized Shannon entropy \rlb{Ssym0} and the symmetrized von Neumann entropy \rlb{Ssym} have deep physical and mathematical meanings, which should be further explored to shed light on universal aspects of nonequilibrium steady states.

It is indeed suggestive to rewrite the symmetrized von Neumann entropy \rlb{Ssym} as
\eq
\Ssym[\rho]=S_\mathrm{vN}[\rho]+D[\rho|\tilde{\rho}]
\lb{Ssym2}
\en
with the relative entropy defined by $D[\rho|\sigma]:=\Tr[\rho(\log\rho-\log\sigma)]$.
Since the relative entropy can be interpreted as the ``distance'' between the two states, the second term $D[\rho|\tilde{\rho}]$ is a qualitative measure of the amount of time-reversal symmetry breaking in the state $\rho$.
We wish to learn if this observation leads us to any deeper understanding of thermodynamics of NESS.

Since we regard the present study as a beginning rather than a goal, there are so many important issues that must be studied in the future.

One thing we wish to know is whether one can show a quantum version of the ``nonlinear nonequilibrium'' improvement of the extended Clausius relation discovered in \cite{KNST2}.
If the relation is really robust and universal, we belive that it should be recovered in the (more general) quantum setting.
For the moment we have no idea about how to derive the relation in quantum systems.
It remains to be understood whether this difficulty is only  technical or  essential.

We also note that, in the present realization of heat baths, we always need to ``refresh'' the baths whenever we change the parameters of the system.
See \rlb{qtot}, footnote~\ref{fn:baths}, and related discussions.
It is quite interesting and important to investigate physically more natural designs of  heat bath.

Last but not least, it is most desirable to make a nontrivial theoretical prediction that can be directly tested by experiments.
A promising approach may be to extend the present approach to the problem of mesoscopic conductors studied, e.g., in \cite{Nakamuraetal}.

\appendix
\section{Representations of NESS}
In this Appendix, we present two expressions which represent NESS in terms of expectation values of time-dependent quantities.
One is a variant of the Zubarev-McLennan representation, and the other is a linear response formula.
We do not claim that these are new, but we were not able to find these expressions explicitly mentioned in the literature.
Although these representations are not used in the present work, we believe that they are interesting enough to be discussed.

\subsubsection*{Zubarev-McLennan representation}
Suppose that the Hilbert space $\HiS$ of the system has a finite dimension $\calN$.
Let $\varphi\in\HiS$ be an arbitrary pure state of the system.
Then by setting $\rho_1$ as the projection operator onto $\TR\varphi$, and $\rho_0=\bsone_\mathrm{S}/\calN$, the Zubarev-McLennan representation \rlb{Zubarev} becomes
\eq
\sbkt{\varphi,\rhoss\,\varphi}
=
 {\calN }^{-1}  \Bbkt{e^{\betavec\cdot(\Ejvec-\Ejvecp)}}_{\TR\varphi\,\to\,\mathrm{ss}} \,
\lb{Zubarev3}
\en
We have defined the new average
\eq
\bbkt{f(\jvec,\jvec')}_{\rho\,\to\,\mathrm{ss}}:=
\sum_{\jvec,\jvec'}f(\jvec,\jvec')\,
\Tr[\,(\bsone_\mathrm{S}\otimes P_{\jvec'})\,U\,(\rho\otimes P_{\jvec})\,U^\dagger\,]
\,\frac{e^{-\betavec\cdot\Ejvec}}{Z(\betavec)},
\lb{fav2}
\en
which is similar to the average \rlb{fav}, but has no conditioning on the final state.
One can naturally interpret it as the average in a process where the system starts from $\rho$ and relaxes into the NESS.

The expression \rlb{Zubarev3} may not be useful as it is, but is of interest since it is an exact representation of the NESS in terms of the average of the quantity $e^{\betavec\cdot(\Ejvec-\Ejvecp)}$, which is related to the total entropy production in the baths.

Although \rlb{Zubarev3} is a representation of a diagonal element of the density matrix $\rhoss$, it also yields information about off-diagonal elements.
Let $\psi$ and $\xi$ be arbitrary pure states which are orthogonal with each other.
Then one easily verifies that
\eq
\sbkt{\psi,\rhoss\,\xi}=
\sbkt{\varphi_\mathrm{r},\rhoss\,\varphi_\mathrm{r}}
-i\sbkt{\varphi_\mathrm{i},\rhoss\,\varphi_\mathrm{i}}
-\frac{e^{-i\pi/4}}{\sqrt{2}}
\bigl\{
\sbkt{\psi,\rhoss\,\psi}+\sbkt{\xi,\rhoss\,\xi}
\bigr\},
\lb{offd}
\en
where
\eq
\varphi_\mathrm{r}=\frac{1}{\sqrt{2}}(\psi+\xi),\quad
\varphi_\mathrm{i}=\frac{1}{\sqrt{2}}(\psi+i\xi).
\en
Since the right-hand side of \rlb{offd} only involves diagonal elements, one can use \rlb{Zubarev3} to represent $\sbkt{\psi,\rhoss\,\xi}$.

\subsubsection*{Remark on the linear response formula}
The reader might notice that the representation \rlb{preLR} is almost the linear response formula.
Let us make this connection explicit.

Although one can start from \rlb{preLR} and directly derive the final result \rlb{LR}, we take a slightly different route.
Observe that \rlb{logprp} with \rlb{bEF} implies
\eq
\log\sbkt{\varphi,\rhosst\,\varphi}=
\beta(\varepsilon_i-\varepsilon)+
\sbkt{\mathcal{F}_{\jvec,\jvec'}}_{\varphi\to\psi_i}
+\log\bbkt{\psi_i,\rhoss\,\psi_i}+O^2.
\lb{preLR2}
\en
From \rlb{Fdef} and \rlb{encon}, one has
\eq
\mathcal{F}_{\jvec,\jvec'}=
\sum_{\nu=1}^n(\beta_\nu-\beta)(E_{j_\nu}^{(\nu)}-E_{j'_\nu}^{(\nu)})
+O(\eta)+O(\gamma).
\lb{F2}
\en
Multiplying \rlb{preLR2} by $\bbkt{\psi_i,\rhoss\,\psi_i}$ and summing  over all $i$ as we did below \rlb{long}, we get
\eq
\log\sbkt{\varphi,\rhosst\,\varphi}=-\beta\varepsilon
+\sum_{\nu=1}^n(\beta_\nu-\beta)
\Tr[\,Q^{(\nu)}\,(P_{\varphi}\otimes\rhoBb)\,]
+\beta\bar{\varepsilon}-S+O(\eta)+O(\gamma)+O^2,
\en
where $\bar{\varepsilon}:=\sum_i\varepsilon_i\bbkt{\psi_i,\rhoss\,\psi_i}=\TrS[\HS\rhoss]$.

Let us now set $\varphi=\psi_j$, which means $\eta=0$.
We also take the weak coupling limit where $\gamma$ becomes negligible.
Finally we replace the initial state of the baths $\rhoBb$ with the equilibrium initial state $\rho_{\mathrm{B},(\beta,\ldots,\beta)}$ where all the baths have the same inverse temperature.
Since $\beta_\nu-\beta=\oeo$, this replacement yields an error term of $\oes$, and one finally gets
\eq
\log\sbkt{\psi_j,\rhosst\,\psi_j}=-\beta\varepsilon_j
+\sum_{\nu=1}^n(\beta_\nu-\beta)
\Tr[\,Q^{(\nu)}\,
(P_{\psi_j}\otimes\rho_{\mathrm{B},(\beta,\ldots,\beta)})\,]
+\beta\bar{\varepsilon}-S+\oes.
\lb{LR}
\en
Note that here the first order correction to the canonical ensemble is represented in terms of the time-dependent correlation function in the equilibrium ensemble.
This is nothing but a linear response representation of the NESS \cite{LR}.
We remark that we are only able to derive the representation \rlb{LR} for diagonal matrix elements of $\rhoss$ for an eigenstate $\psi_i$ of $\HS$.

\bigskip
{\small 
It is a pleasure to thank Tohru Koma, Teruhisa Komatsu, Naoko Nakagawa, and Shin-ichi Sasa for useful discussions.
}



\begin{thebibliography}{10}

\bibitem{LR}
R. Kubo, M. Toda and N. Hashitsume, 
{\em Statistical Physics II: Nonequilibrium Statistical Mechanics}\/,
(Springer-Verlag, Berlin, 1991).

\bibitem{Zubarev}
D. N. Zubarev, 
{\em Nonequlibrium Statistical Thermodynamics}\/, (Consultants Bureau, New York, 1974).

\bibitem{McLennan}
J. A. McLennan, 
{\em Introduction to Nonequilibrium Statistical Mechanics}\/,
(Prentice Hall, Englewood Cliffs, NJ, 1990).

\bibitem{KG}
K. Kawasaki and J. D. Gunton,
{\em Theory of Nonlinear Transport Processes: Nonlinear Shear Viscosity and Normal Stress Effects}\/,
Phys. Rev. A {\bf 8} 2048--2064 (1973).
 
\bibitem{Evans}
D. J. Evans, E. G. D. Cohen and G. P. Morris, 
{\em Probability of second law violations in shearing steady states}\/,
Phys. Rev. Lett. {\bf 71} 2401 (1993).

\bibitem{GC}
G. Gallavotti and E. G. D. Cohen, 
{\em Dynamical Ensembles in Nonequilibrium Statistical Mechanics}\/,
Phys. Rev. Lett. {\bf 74}, 2694 (1995).

\bibitem{Crooks}
G. E. Crooks, 
{\em Entropy production fluctuation theorem and the
nonequilibrium work relation for free energy differences}\/,
Phys. Rev. E {\bf 61}, 2361--2366 (2000), {\tt cond-mat/9901352}. 

\bibitem{Kurchan}
J. Kurchan, 
{\em A quantum fluctuation theorem}\/, preprint (2000), 
{\tt cond-mat/0007360}.

\bibitem{Tasaki}
H. Tasaki, 
{\em Jarzynski relations for quantum systems and some applications}\/,
unpublished note (2000), {\tt cond-mat/0009244}. 

\bibitem{hanggi}
Michele Campisi, Peter Hanggi and Peter Talkner
{\em Colloquium. Quantum Fluctuation Relations: Foundations and Applications}\/,
Reviews of Modern Physics, to be published,  {\tt arXiv:1012.2268}.

\bibitem{LD1}
B. Derrida J. L. Lebowitz and E. R. Speer,
{Large deviation of the density profile in the steady
state of the open symmetric simple exclusion process}\/,
J. Stat. Phys. {\bf 107}, 599--634 (2002), {\tt cond-mat/0109346}.  

\bibitem{LD2}
K. Saito and A. Dhar,
{\em Fluctuation theorem in Quantum Heat Conduction}\/,
Phys. Rev. Lett. {\bf 99}, 180601 (2007), {\tt cond-mat/0703777}.  


\bibitem{KNST1}
T. S. Komatsu, N. Nakagawa, S. Sasa and H. Tasaki,
{\em Steady State Thermodynamics for Heat Conduction --- Microscopic Derivation}\/,
Phys. Rev. Lett. {\bf 100}, 230602 (2008), {\tt arXiv:0711.0246}.

\bibitem{KNST2}
T. S. Komatsu, N. Nakagawa, S. Sasa and H. Tasaki,
{\em Entropy and Nonlinear Nonequilibrium Thermodynamic Relation for Heat Conducting Steady States}\/,
J. Stat. Phys. {\bf 142}, 127--153,  {\tt arXiv:1009.0970}.

\bibitem{OP}
Y. Oono and M. Paniconi,
{\em Steady state thermodynamics}\/,
Prog. Theor. Phys. Suppl.
\textbf{130}, 29-44 (1998).

\bibitem{Landauer}
R. Landauer, {\em $dQ= TdS$ far from equilibrium}\/, Phys. Rev. A18, 255-266 (1978).

\bibitem{Ruelle}
D. Ruelle,
{\em Extending the definition of entropy to nonequilibrium steady states}\/,
Proc. Natl. Acad. Sci. U.S.A.  {\bf 100}, 3054--3058 (2003), 
{\tt arXiv:cond-mat/0303156}.

\bibitem{Nakamuraetal}
S. Nakamura, Y. Yamauchi, M. Hashisaka, K. Chida, K. Kobayashi, T. Ono, R. Leturcq, K. Ensslin, K. Saito, Y. Utsumi, and A. C. Gossard,
{\em Fluctuation Theorem and Microreversibility in a Quantum Coherent Conductor}\/, 
Phys. Rev. B {\bf 83}, 155431 (2011)
, {\tt arXiv:1101.5850}.



\end{thebibliography}
\end{document}